\documentclass[aps, prl, amsmath, superscriptaddress,preprintnumbers,twocolumn]{revtex4-2}
\usepackage{graphicx}
\usepackage{dcolumn}
\usepackage{bm}
\usepackage{amsmath}
\usepackage{subfigure}
\usepackage{amsfonts}
\usepackage[svgnames]{xcolor}
\usepackage{appendix}
\usepackage{multirow}
\usepackage{booktabs}
\usepackage{tabularx}
\usepackage[colorlinks,
            linkcolor=blue,
			filecolor=blue,
			urlcolor= blue,
			citecolor=blue,
            ]{hyperref}
\usepackage{amsthm}
\usepackage{setspace}
\usepackage[T1]{fontenc}
\usepackage{extarrows}
\usepackage{lineno}

\usepackage[percent]{overpic}
\usepackage{dsfont}
\usepackage{floatrow}
\usepackage{physics}

\newcommand{\ee}{{e}}
\newcommand{\nm}{{N}}
\renewcommand{\a}{{\vartheta}}
\newcommand{\s}{{\alpha}}
\newcommand{\m}{{\mu}}
\newcommand{\ii}{{i}}
\newcommand{\nn}{{\hat{n}}}

\newcommand{\M}[0]{{\mathcal{M}}}

\newcommand{\bz}[0]{\bar{Z}}
\newcommand{\bx}[0]{\bar{X}}

\begin{document}

\title{General Approach to Error Detection of Bosonic Codes via Phase Estimation}

\author{Yuan-De Jin}
\affiliation{State Key Laboratory of Semiconductor Physics and Chip Technologies, Institute of Semiconductors, Chinese Academy of Sciences, Beijing 100083, China}
\affiliation{Center of Materials Science and Opto-Electronic Technology, University of Chinese Academy of Sciences, Beijing 100049, China}

\author{Shi-Yu Zhang}
\affiliation{Department of Physics, University of Michigan, Ann Arbor, Michigan 48109, USA}

\author{Ulrik L. Andersen}
\affiliation{Department of Physics, Center for Macroscopic Quantum States (bigQ), Technical University of Denmark,
Fysikvej, 2800 Kgs., Lyngby, Denmark}

\author{Wen-Long Ma}
\email{wenlongma@semi.ac.cn}
\affiliation{State Key Laboratory of Semiconductor Physics and Chip Technologies, Institute of Semiconductors, Chinese Academy of Sciences, Beijing 100083, China}
\affiliation{Center of Materials Science and Opto-Electronic Technology, University of Chinese Academy of Sciences, Beijing 100049, China}
\date{\today }
\begin{abstract}
We present a general approach to error detection of bosonic quantum error-correction codes via an adaptive quantum phase estimation algorithm assisted by a single ancilla qubit. The approach is applicable to a broad class of bosonic codes whose error syndromes are described by symmetry or stabilizer operators, including the rotation-symmetric codes and Gottesman-Kitaev-Preskill (GKP) codes. The detection precision scales inversely with the total evolution time and thus reaches the Heisenberg limit. We numerically demonstrate the approach for several examples, such as detecting bosonic excitation loss errors in high-order cat or binomial codes and displacement errors in finite-energy GKP codes. We also extend the approach to efficiently generate arbitrary Fock states. Our schemes are feasible in present-day experiments. 


\end{abstract}

\maketitle

\textit{Introduction.} 
Qubits are inevitably coupled to their surrounding environments, leading to quantum dissipation and decoherence \cite{Zurek2003,Chirolli2008b,Schlosshauer2019e,Qiu2024a}. Quantum error correction (QEC) is a systematic strategy to protect quantum information against environmental noise and control errors and is indispensable to large-scale quantum information processing \cite{Shor1995,Cochrane1999,Fowler2012,Devitt2013,Roffe2019,Hastrup2022}.
Bosonic QEC codes are typically implemented in circuit quantum electrodynamics (cQED) \cite{Blais2020,Joshi2021,Ma2021,Cai2021a}, trapped ions \cite{Fluhmann2019,DeNeeve2022} and quantum acoustic systems \cite{Chu2017,VonLupke2022}. Due to the hardware efficiency compared to conventional QEC codes based on multiple qubits \cite{Albert2018,Cai2021a,Albert2022}, bosonic codes have first reached the break-even point \cite{Ofek2016,Hu2019,Ni2023,Sivak2023}, beyond which the coherence time of the logical qubit exceeds those of the constituent physical qubits. According to the code symmetry, current single-mode bosonic codes can be classified into rotation-symmetric codes \cite{Grimsmo2020} (e.g., cat codes \cite{Cochrane1999, Leghtas2013} and binomial codes \cite{Michael2016}) and the translation-symmetric codes \cite{Grimsmo2021} (e.g., Gottesman-Kitaev-Preskill (GKP) codes \cite{Gottesman2001}). The former can correct bosonic loss and dephasing errors, while the latter can 
correct displacement errors. To implement QEC, it is crucial to design efficient quantum non-demolition (QND) measurement protocols to detect the errors. 

For the simplest rotation-symmetric codes, the error syndrome for a single bosonic loss error is the even-odd number parity of bosonic excitations, which can be efficiently extracted by Ramsey interferometry measurement (RIM) \cite{Vlastakis2013, Sun2014,Hu2019} or frequency comb control \cite{Ni2023} of an ancilla qubit. For general rotation-symmetric codes, the error syndromes are the modular numbers of bosonic excitations \cite{Michael2016}. Current proposals to detect such modular numbers often need a higher-dimensional ancilla, such as a qudit to perform $\mathbb{Z}_d$ measurements \cite{Li2017} or an oscillator to perform controlled-rotation gates \cite{Grimsmo2020}. However, the engineering of a complex ancilla is highly demanding for most platforms, hindering the realization of bosonic QEC with high-order rotation-symmetric codes to correct more bosonic errors. On the other hand, it has been proposed to use quantum phase estimation (QPE) algorithms to prepare GKP states with an ancilla qubit \cite{Terhal2016,Shi2019}. Moreover, assisted by quantum optimal control, an ancilla qubit dispersively coupled to a single bosonic mode can detect the binary bits of photon number \cite{Wang2020} or prepare large Fock states \cite{Deng2024}. However, it is unknown whether these approaches can apply to general bosonic codes. Therefore, there is an urgent need of a general approach for generation and error detection of general bosonic codes, guiding the implementation of more efficient bosonic QEC within current experimental limitations.



In this paper, by building the connection between error detection of bosonic QEC codes with the adaptive QPE algorithm, we provide a unified framework for preparation and error detection of various single-mode bosonic codes, whose error syndromes are described by some symmetry or stabilizer operators. We demonstrate that the detection of modular number for rotation-symmetric codes and displacement for translation-symmetric codes can both be incorporated into this framework. Then we design an efficient protocol for error detection of cat, binomial and GKP codes based on sequential adaptive RIMs on a single ancilla qubit, which can theoretically achieve the Heisenberge limit (HL) in detection precision and is feasible to implement in practical systems. We further extend this protocol to efficiently prepare arbitrary Fock states.

\begin{figure*}
    \centering
    \includegraphics[width=18cm]{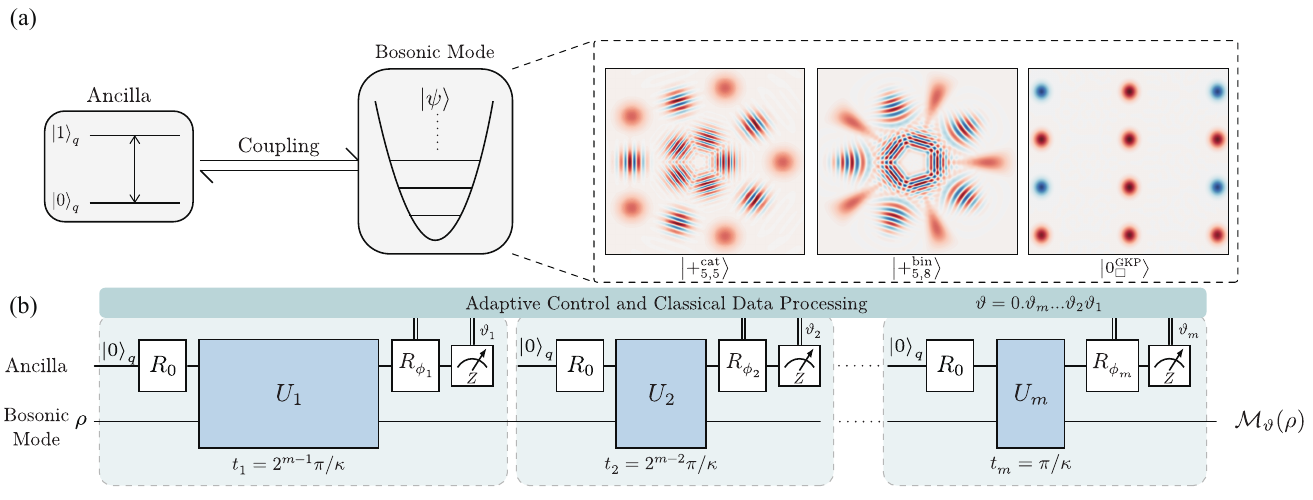}
    \caption{Schematic of error detection of bosonic codes with adaptive QPE. (a) A single bosonic mode coupled to an ancilla qubit.  (b) The quantum circuit of adaptive QPE. Our protocol is applicable to both rotation-symmetric and GKP codes.} 
    \label{fig1}
\end{figure*}

\textit{Error detection of bosonic codes as a QPE problem}. The QPE algorithm is crucial for estimating the eigenvalues of a $d$-dimensional unitary operator \cite{Kitaev1995,OBrien2019,Ding2023d,Somma2019,Jin2025}. For a unitary operator $\ee^{i V\tau}$ generated by a Hermitian operator $V=\sum_{k=1}^{d} v_k |k\rangle\langle k|$ with $\{v_k\}_{k=1}^d$ and $\{|k\rangle\}_{k=1}^d$ being the set of eigenvalues and orthonormal basis, respectively. The objective of QPE is to estimate the set of eigenvalues $\{\ee^{i v_k\tau}\}_{k=1}^d$. 
The main finding of this paper is to provide a rigorous approach based on sequential quantum channels to recast the error detection of bosonic codes as the problem of QPE of multiple eigenvalues. 

Suppose a single bosonic mode is coupled to an ancilla qubit with the Hamiltonian [Fig. \ref{fig1}{\color{blue}(a)}] 
\begin{equation}
  H=\sigma_q^z\otimes V,
\end{equation}
where $\sigma_q^i$ $(i=x,y,z)$ is the Pauli operator of the ancilla, and $V$ is a discrete or continuous observable of the bosonic system corresponding to the symmetry operator of a bosonic QEC code. For a discrete symmetry operator,  $V=\sum_{k=1}^\infty v_k |k\rangle\langle k|$ has infinite eigenvalues. However, $\ee^{iV\tau}$ can have a finite number of nondegenerate eigenvalues by selecting a proper $\tau$, i.e., $\ee^{iV\tau}=\sum_{l} \ee^{iv_l\tau}\Pi_l$ with $\Pi_l$ being the projection to the code or error subspace. For a continuous symmetry operator (corresponding to the translation-symmetric code), the decomposition of $V$ is similar except that the summations above are replaced by integration.

Our QPE scheme is based on sequential adaptive RIMs [Fig. \ref{fig1}{\color{blue}(b)}]. The $i$th RIM starts with an initial rotation $R_0$, with $R_{\phi}=e^{-i(\cos\phi \sigma_q^x+\sin \phi \sigma_q^y)\pi/4}$ denoting the $\pi/2$ rotation of the ancilla, followed by a free evolution of the composite system $U_i=e^{-iH t_i}=\sum_{\s=0,1}\ket{\s}_q\bra{\s}\otimes U_{\s,i}$ with $\ket{\alpha}_q\,(\alpha\in\{0,1\})$ denoting the ancilla state, $U_{\s,i}=e^{(-1)^{\s+1}iVt_i}$ and $t_i=2^{m-i}\pi/\kappa$ ($\kappa$ is a constant defined below), and finally a second ancilla rotation $R_{\phi}$ with $\phi_i=\pi-2\pi0.0\a_{i-1}...\a_2\a_1$ conditioned on previous measurement outcomes $\{\a_1, \a_2,...,\a_{i-1}\}$ ($\a_j\in\{0,1\}$) (see Fig. \ref{fig1}). 
The measurement time and phase feedback in QPE can be understood if all the eigenvalues can be approximated by $m$-bit binary numbers. The outcome $\a_i$ of the $i$-th measurement gives the estimation of the $(m-i+1)$-th bit in the binary expansion. Thus, a single binary number $\a=0.\a_m...\a_2\a_1$ determined by all the measurement results $\{\a_1,\a_2,...\a_m\}$ represents an approximation of $v_k/\kappa$. 


We can exactly solve the dynamics of bosonic modes in adaptive QPE with the framework of quantum channels \cite{Jin2025}. Each RIM cycle induces a quantum channel on the target system \footnote{See Supplemental Material including Ref. \cite{Kraus1983,Ma2023c,Jin2025,Royer2020,Mostafazadeh2005,Scolarici2006,Ohlsson2021,Grimsmo2020,Terhal2016,Hastrup2021a,Didier2015} at \url{} for more details of the framework for error detection with QPE, numerical simulations of bosonic code and arbitrary Fock state preparation and performance evaluation.}
\begin{equation}
  \Phi_i(\rho)={\rm Tr}_q[U_i(t_i)(\rho_q\otimes \rho)U_i^{\dagger}(t_i)]=\sum_{\a_i=0}^1\mathcal{M}_{\a_i}(\rho),
\end{equation}
where $\rho_q=R_{0}|\psi\rangle_q\langle \psi|R_{0}^\dagger$, $\rho$ is the initial state of the bosonic system, ${\rm Tr}_q[\cdot]$ denotes the partial trace over the ancilla, and $\mathcal{M}_{\a_i}(\cdot)=M_{\a_i} (\cdot) M_{\a_i}^{\dagger}$ is a superoperator with the Kraus operator $M_{\a_i}=[U_{0,i}-(-1)^{\a_i}e^{i\phi_i}U_{1,i}]/2$. 

After $m$ measurements, sequential quantum channels can be decomposed into quantum trajectories as
$
   \Phi_m \cdots \Phi_2\Phi_1(\rho) =\sum_{\a_1,\cdots,\a_m}\M_\a(\rho)
$, where $\M_\a(\cdot):=\M_{\a_m}\cdots \M_{\a_2}\M_{\a_1}(\cdot)$.
For the outcome $\a$ of sequential RIMs, the bosonic system is steered to 
$\rho'={\M_\a(\rho)}/{\Tr[\mathcal{M}_\a(\rho)]}$ with
\begin{equation}\label{M}
\begin{aligned}
	    \M_\a(\rho)&=\sum_{k,l}(-1)^{b_{k,l}}\sqrt{F_{2^m}\left(\a-\frac{v_k}{\kappa}\right)
	    F_{2^m}\left(\a-\frac{v_l}{\kappa}\right)}\Pi_k \rho \Pi_l,
\end{aligned}
\end{equation}
where $F_{N}(x):=\left[\frac{\sin(N\pi x)}{N\sin(\pi x)}\right]^2$ is a distorted Fej\'{e}r kernel, and $b_{k,l}=\lfloor \xi_k\rfloor+\lfloor 2^{m}\xi_k\rfloor+\lfloor \xi_l\rfloor+\lfloor 2^{m}\xi_l\rfloor$ with $\xi_k=\a-v_k/\kappa$ and  $\lfloor~\rfloor$ being the floor function. 
    
For the symmetry operator $V$, we can detect the error syndromes of bosonic QEC codes through estimation of eigenvalues of $\ee^{iVt_m}$, simultaneously preparing the bosonic mode in the corresponding code or error subspace. The probability to obtain the outcome $\a$ is 
$p(\a)=\Tr[\mathcal{M}_\a(\rho)]= \sum_{k} {\rm Tr}(\rho \Pi_k)F_{2^{m}}(\a-v_k/\kappa)$, so the measurement statistics of $\a$ shows a finite (or continuous) set of Fej\'{e}r peaks around the eigenvalues of $V$ with weight $\Tr(\rho \Pi_k)$. Since the distribution width of the Fej\'er kernel $F_{2^m}(\cdot)$ is proportional to $2^{-m}$, the ideal estimation accuracy of this scheme can reach the HL, i.e. $\delta\sim t_{\rm tot}^{-1}$ 
with $\delta$ being the estimation error and $t_{\rm tot}=\sum_{i=1}^m t_i=(2^m-1)\pi/\kappa$ the total evolution time. 



\textit{Error detection of rotation-symmetric codes and arbitrary Fock state generation}. A bosonic code with $N$-fold rotational symmetry means that any logical code state is an eigenstate of the rotation operator $R_N=e^{\ii(2\pi/N)\nn}$ with the eigenvalue $+$1 \cite{Grimsmo2020}, where $\nn=a^\dagger a$ is the boson number operator with $a^\dagger(a)$ being creation (annihilation) operator. We can define $\bar{Z}=\sqrt{R_N}=R_{2N}=e^{\ii(\pi/N)\nn}$ as logical $\bar Z$ operator since logical states are eigenstates of $\bz$ with eigenvalue $\pm 1$. Then the rotation-symmetric codes have a unified form in the Fock space:
$ \ket{\mu_N}=\sum_{k=0}^\infty f_{(2k+\m)N}\ket{(2k+\m)N},$
where $\m\in\{0, 1\}$, and $f_{kN}$ is a code-dependent superposition coefficient. 

Then we can perform error detection by dispersively coupling the bosonic mode to an ancilla qubit. The coupling Hamiltonian is 
\begin{equation}\label{Disp}
  H=-\chi|1\rangle_e\langle 1|\otimes \nn,
\end{equation}
in which the number operator can be decomposed as $ \nn = \sum_{l=0}^{N-1}\sum_{k=0}^\infty (l+kN)\ket{l+kN}\bra{l+kN}$. For $N$-fold rotational code, we choose the evolution time $t_i=2^{m-i+1}\pi/(\chi N)$, the measurement statistics becomes \cite{Note1}
\begin{equation}
    p(\a)=\sum_{l=0}^{N-1} \Tr(\rho\Pi_{N}^l)F_{2^m}\left(\a-\frac{l}{N}\right),\label{Eq:p(a)}
\end{equation}
where $\Pi_N^l=\sum_{k=0}^\infty \ket{l+kN}\bra{l+kN}$.
So we can detect the bosonic errors from the measurement statistics of $\a$. For a bosonic state in the code subspace, the distribution of $\a$ would exhibit a single peak around 0 since $\Tr(\rho\Pi_{N}^l)=\delta_{l0}$, while for a state in the error subspace suffering $l$ ($<N$) bosonic excitation loss (gain) error, we expect to observe a peak around $1-l/N$ ($l/N$). However, when the system loses or gains $l$ ($\geq N$) excitations exceeding the code's correction range, the measured error syndrome is returned modulo $N$ as $l \mod N$, which is thus indinguishable with that of the correctable error. 

The above scheme can be extended to prepare arbitrary Fock states of a bosonic mode. For an initial bosonic state with the maximum excitation number $M$ and the QPE sequence that can detect a maximum modular number $N$, we distinguish two cases: (i) if $N>M$, we can directly prepare a Fock state $\ket{n}$ ($n\leq M<N$) through the QPE sequence with the modular number $N$; (ii) if $N\leq M$, we can combine several QPE sequences with different modular numbers $\{\nm_i\}$ to prepare large Fock states ($N\leq n\leq M$). According to the Chinese remainder theorem, an integer can be uniquely determined if its remainders when divided by several pairwise coprime integers are known. Then by choosing some pairwise coprime $\{\nm_i\}_{i=1}^r$ ($N_i<N$), the maximum detectable excitation number is $M=\prod_{i=1}^r \nm_i$ \cite{Note1}.

\begin{figure*}
    \centering
    \includegraphics[width=18cm]{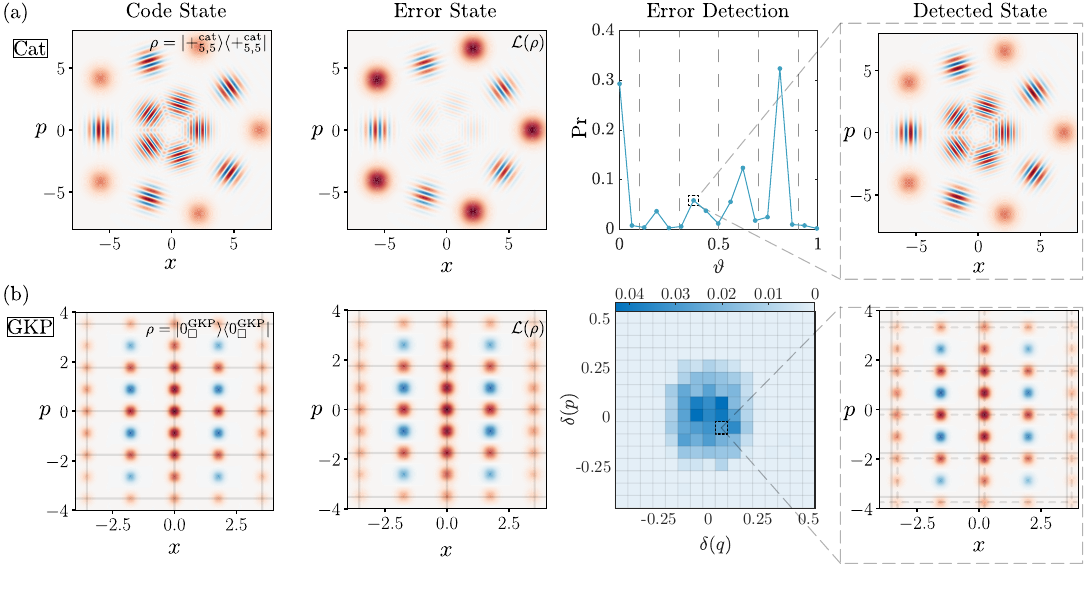}
    \caption{Error detection via adaptive QPE for the (a) cat code and (b) GKP code. The error states are produced by the lossy bosonic channel, and the probability distributions of error detection are shown along with typical detected states. For GKP states, we show the distribution of $\delta(x,p)=\a(x,p)-\operatorname{round}[\a(x,p)]$ to represent the actual displacement, and the original (detected) GKP lattices are labeled by solid (dashed) grey lines. Here we take a truncated Fock space with dimension 100, 701 for the cat and GKP code respectively, and the bosonic loss probability as $\gamma=0.03$. }
    
    \label{FigMC}
\end{figure*}

\textit{Error detection of GKP codes}. GKP codes have discrete translation symmetry. The logical code with such symmetry should be simultaneously +1 eigenstate of the displacement operators $S_Z=D(2\alpha)$ and $S_X=D(2\beta)$ ($S_Z$ and $S_X$ are also called stabilizer), where $D(\alpha)=e^{\alpha a^\dagger-\alpha^* a}$, then $\bz=\sqrt{D(2\alpha)}=D(\alpha)$ and $\bx=D(\beta)$ can be defined. In order to suffice the anti-commutation rule of Pauli operators, i.e. $\{\bz,\bx\}=0$, the relation $\alpha^*\beta-\alpha\beta^*=\ii\pi$ should hold since $D(\beta)D(\alpha)=e^{\alpha^*\beta-\alpha\beta^*}D(\beta)D(\alpha)$. We further define two generalized quadratures $Q=\ii (\beta^* a-\beta a^\dagger)/\sqrt{\pi}$ and $P=-\ii(\alpha^* a -\alpha a^\dagger)/\sqrt{\pi}$, then the canonical commutator $[Q,P]=i$ is satisfied. Besides, we have $\bx=e^{-\ii\sqrt{\pi}P}$ and $\bz=e^{\ii\sqrt{\pi}Q}$, then the logical code can be defined naturally with this form,
$\ket{\m^{\rm GKP}}=\sum_{k=-\infty}^\infty \ket{(2k+\m)\sqrt{\pi}}_Q=\sum_{k=-\infty}^{\infty} e^{-ik\pi\m}|k\pi\rangle_P,$
with $Q\ket{x}_Q=x\ket{x}_Q$ and $P\ket{p}_P=p\ket{p}_P$.

We consider the GKP code $\ket{\m^{\rm GKP}_{\square}}$ in the square lattice with $\alpha=\sqrt{\frac{\pi}{2}}$ and $\beta=\ii\sqrt{\frac{\pi}{2}}$, then the generalized quadratures are the usual position and momentum quadratures, i.e., $Q=(a+a^\dagger)/\sqrt{2}$ and $P=-i(a-a^\dagger)/\sqrt{2}$. Thus we can detect the displacement errors along the $Q$ and $P$ axes respectively, by sequentially choosing the coupling Hamiltonian between the ancilla and bosonic mode 
\begin{equation}
\begin{aligned}
      H_\theta&=g\sigma_q^z\otimes (a e^{-i\theta}+a^\dagger e^{i\theta}),
\end{aligned}
 \end{equation}
with $\theta=0,\pi/2$ for quadratures $Q$ and $P$.
By taking the evolution time $t_i=2^{m-i}\pi/(g\sqrt{2\pi})$, the measurement statistics for the quadrature $Q$ becomes \cite{Note1}
\begin{equation}
    p[\a(x)]=\int_{-\sqrt{\pi}/2}^{\sqrt{\pi}/2} \Tr(\rho \Pi_{Q}^x)F_{2^m}\left(\a(x)- \frac{x}{\sqrt{\pi}}\right) \dd x,
\end{equation}
with $\Pi_{Q}^x=\sum_{k=-\infty}^\infty|x+k\sqrt{\pi}\rangle_Q\langle x+k\sqrt{\pi}|$. 
For the outcome $\vartheta(x)=0.\vartheta_{x,1}\vartheta_{x,2}...\vartheta_{x,m}$, the displacement error is expected to be $\Delta x=\vartheta(x)\sqrt{\pi}(\operatorname{mod} \sqrt{\pi})$. Similarly, for the outcome $\vartheta(p)$, the displacement error is approximately $\Delta p=\vartheta(p)\sqrt{\pi}(\operatorname{mod} \sqrt{\pi})$, with $\vartheta(p)=0.\vartheta_{p,1}\vartheta_{p,2}...\vartheta_{p,m} $.


\textit{Examples.} Below we numerically demonstrate our scheme in several examples, including the cat codes, binomial codes and GKP codes. For all bosonic codes, we consider the errors induced by the Lindblad dissipation $\rho'=\mathcal{L}(\rho)=\exp{\chi \mathcal{D}(\rho)}$ with $\mathcal{D}(\rho)=a\rho a^\dagger-\frac{1}{2}\{\nn,\rho\}$, which can also be decomposed to a lossy bosonic channel, i.e., $\mathcal{L}(\rho)=\sum_{n=0}^\infty E_k \rho E_k^\dagger$, with $E_k=(1/\sqrt{k!})\gamma^{k/2}(1-\gamma)^{a^{\dagger}a/2}a^k$ being the Kraus operator associated with losing $k$ excitations and $\gamma=1-e^{-\chi}$ the loss probability of each excitation \cite{Albert2018}.

First we consider a cat code with $N=5$ and $\alpha=5$. The logical qubit of cat code is $|\mu_{N,\alpha}^{\rm cat}\rangle=\sum_{m=0}^{2 N-1}(-1)^{\mu m} |e^{i(m \pi / N) }\alpha\rangle$. We perform Monte Carlo simulations to show the error detection for the state $\ket{+_{N,\alpha}^{\rm cat}}=(\ket{0_{N,\alpha}^{\rm cat}}+\ket{1_{N,\alpha}^{\rm cat}})/\sqrt{2}$. When the code state experiences a lossy process, the error detection distribution shows multiple
peaks around $l/5$ ($l=1, ..., 4$) [Fig. \ref{FigMC}{\color{blue}(a)}]. For the samples with outcomes near $l/N$, it can be deduced that $(N-l) \mod N$ excitations are lost, and the state is also projected to such an error subspace, i.e. $\rho_{\rm error}\propto (a)^{N-l}\ket{+_{N,\alpha}^{\rm cat}}\bra{+_{N,\alpha}^{\rm cat}} (a^\dagger)^{N-l}$, so that a further recovery operation can be applied to correct the error. 
We also confirm the HL in detection precision (see \cite{Note1} for Monte Carlo simulations). Moreover, the errors of binomial codes can also be detected with the same approach \cite{Note1}. 

Then we consider the GKP code. Due to the non-normalizability of ideal GKP code, we use the approximate GKP code \cite{Grimsmo2021,Hastrup2021,Royer2020,Menicucci2014}. With a parameter $\Delta$,  
 $\ket{\tilde\m^{\rm GKP}_{\square}}\propto e^{-\Delta^2a^\dagger a} \ket{\m^{\rm GKP}_{\square}}$
is an approximation of $ \ket{\m^{\rm GKP}_{\square}}$. Then $\ket{\tilde \m^{\rm GKP}_{\square}}$ becomes the eigenstate of $\bz_\Delta=e^{i\sqrt{\pi}Q_\Delta}$ with $Q_\Delta=\cosh\Delta^2Q+i\sinh\Delta^2P$. In our simulations, we set $\Delta\approx0.251$ producing 12 dB squeezing. From the Monte Carlo simulation and Wigner function shown in Fig. \ref{FigMC}{\color{blue}(b)}, we see the error of lossy channel can be detected, while the specific trajectory shows the specific displacement projected from the initial state corresponding to the measurement outcomes. 

In our QPE scheme, the error-detection infidelity arises from two competing factors. First, when the number of measurements $m$ (total measurement time $t_{\rm tot}$) is small, the QPE scheme suffers from some deduction errors due to the broadening of the Fej\'er kernels. Second, for large $m$, the error detection should be limited by the significant errors of the ancilla and bosonic system in practical systems \cite{Note1}. So as $m$ increases, typically the infidelity will first decrease and then increase, implying that there exists some optimal total measurement time. In simulations for practical cQED systems, the overall error-detection infidelity can reach 0.05 for rotation-symmetric bosonic codes and 0.058 for the GKP code \cite{Note1}. Note that our analysis and simulations assume perfect ancilla state initialization and readout, while realistic readout and initialization errors would affect the fidelity. Moreover, the precision of QPE also relies on well-calibrated controlled-unitary operations, while practical miscalibrations or fluctuations (e.g. qubit frequency drift, higher-order nonlinearities) can further reduce the precision. We expect such errors can be mitigated using quantum optimal control techniques \cite{werschnik2007quantum,liu2017quantum,yuan2023optimal,Deng2024,gautier2025optimal} to design error-robust pulses, allowing the QPE sequence to be executed faster than the time scale of the environmental noise.

\textit{Experimental considerations.} Finally we discuss the feasibility of our scheme in cQED and ion trap systems. cQED devices are already capable of realizing the essential ingredients of our protocol. The number parity read-out for cat and binomial codes has been demonstrated repeatedly, first with Ramsey-type measurements and later in a fault-tolerant setting by real-time tracking of photon jumps~\cite{Sun2014,Ofek2016,Hu2019,Sivak2023}. The controlled-phase (CROT) operations required for QPE 
can be engineered with the dispersive interaction [Eq.~\eqref{Disp}] together with frequency-selective pulses. Rabi-type couplings between a cavity mode and a transmon, and the tunable dispersive couplers used for CROT gates in the multi-cat architecture of Ref.~\cite{Sivak2023} show that high-fidelity controlled rotations on storage modes are implementable. The main practical consideration is the circuit depth: naively, QPE requires sequential applications of $U^{2^k}$, so the gate time grows exponentially with the desired binary precision. For example, consider $m=4$ adaptive rounds and $\chi/2\pi=2$ MHz, yielding a total interaction time $t_{\rm tot} \approx 3.75~\mu{\rm s}$. For the storage lifetimes reported in Ref.~\cite{Sivak2023} ($T_1\sim60\mu$s), today’s hardware comfortably accommodates one or two binary digits, which already detects up to two photon-loss events.

In radio-frequency Paul traps, the harmonic motional mode plays the role of the bosonic oscillator, while internal electronic states serve as ancillas. Jaynes-Cummings and red/blue-sideband interactions furnish the same Rabi coupling identified above, and have supported motional cat and even GKP state preparation~\cite{Fluhmann2019,DeNeeve2022}. State-dependent forces or geometric-phase gates provide a direct route to QPE with precise optical-phase control. Ions therefore possess all primitives needed for a faithful implementation of our QPE-based syndrome extraction. Their exceptionally long coherence (seconds in the motion, minutes in hyperfine qubits) offsets the slower gate rates and mild motional heating that accompany long sequences. In addition, universal phase-space displacements driven by resonant laser pulses allow continuous-valued versions of QPE that are uniquely natural to this architecture.

\textit{Conclusions and outlooks}.
In conclusion, we have developed a QPE algorithm-based framework for constructing general QND measurements of bosonic QEC codes. For rotation-symmetric codes, the preparation and error detection is significantly simplified and optimized, as it can be achieved using dispersive coupling alone without the need for additional driving, while for GKP codes, we demonstrate the feasibility of QPE error detection when discretizing the continuous set of errors. 


Although the adaptive QPE scheme can theoretically achieve the HL in error detection, decoherence of the ancilla and system can degrade this advantage. We expect more elaborate Bayesian or machine-learning-based adaptations \cite{xiao2019continuous,costa2021benchmarking,Yamamoto2024} (e.g., dynamically adjusting the interaction duration or ancilla rotation angle depending on intermediate results) can squeeze out maximal information per ancilla interrogation. Moreover, recent researches have shown that HL can be achieved in noisy systems by using QEC or ancilla entanglement \cite{zhou2018achieving,Zhou2024,Fan2024}. We speculate extending these ideas (e.g., exploring entangling multiple bosonic modes or multiple ancillas) may allow for more rapid and accurate error syndrome detection. Such advanced schemes may be able to reach the HL with fewer resources, and also offer more robustness if decoherence is a limiting factor.  

The adpative QPE framework can also be extended to preparation and error detection of virtually any bosonic codes described by stabilizer or symmetry operators \cite{Albert2019c,Niu2018}, even those involving multiple systems or higher-dimensional eigenspaces constituting the next generation of bosonic codes, such as multimode entangled codes \cite{royer2022encoding,xu2024multimode,lin2023closest}, hybrid concatenation codes \cite{xu2023qubit,putterman2025hardware}, and codes addressing exotic error channels \cite{Leviant2022quantumcapacity,Gravina2023}. 


The research is supported by the National Natural Science Foundation of China (No. 12174379, No. E31Q02BG), the Chinese Academy of Sciences (No. E0SEBB11, No. E27RBB11), the Innovation Program for Quantum Science and Technology (No. 2021ZD0302300) and Chinese Academy of Sciences Project for Young Scientists in Basic Research (YSBR-090). ULA acknowledges support from the Danish National Research Foundation (bigQ, DNRF0142), EU project CLUSTEC (grant agreement no. 101080173), and EU ERC project ClusterQ (grant agreement no. 101055224, ERC-2021-ADG).

\bibliography{Bosonic2.bib}

\onecolumngrid 

\clearpage


\end{document}


\title{ Supplemental Material for "General Approach to Error Detection of Bosonic Codes via Phase Estimation"}

\author{Yuan-De Jin}
\affiliation{State Key Laboratory of Semiconductor Physics and Chip Technologies, Institute of Semiconductors, Chinese Academy of Sciences, Beijing 100083, China}
\affiliation{Center of Materials Science and Opto-Electronic Technology, University of Chinese Academy of Sciences, Beijing 100049, China}

\author{Shi-Yu Zhang}
\affiliation{Department of Physics, University of Michigan, Ann Arbor, Michigan 48109, USA}

\author{Ulrik L. Andersen}
\affiliation{Department of Physics, Center for Macroscopic Quantum States (bigQ), Technical University of Denmark,
Fysikvej, 2800 Kgs., Lyngby, Denmark}

\author{Wen-Long Ma}
\email{wenlongma@semi.ac.cn}
\affiliation{State Key Laboratory of Semiconductor Physics and Chip Technologies, Institute of Semiconductors, Chinese Academy of Sciences, Beijing 100083, China}
\affiliation{Center of Materials Science and Opto-Electronic Technology, University of Chinese Academy of Sciences, Beijing 100049, China}
\date{\today }
\begin{abstract}
	In this Supplementary Material, we first give details about the framework for error detection of bosonic QEC codes using the adaptive QPE algorithm. Then we present extensive numerical simulations demonstrating prepartion of cat, binomial and GKP states, error detection of binomial codes and generation of arbitrary large Fock states. Finally we evaluate the performance of the error detection scheme.
\end{abstract}
\maketitle
\tableofcontents

\section{Details of adaptive quantum phase estimation (QPE)}
For the coupling Hamiltonian 
\begin{equation}
    H=\sigma_q^z\otimes V=\sigma_q^z\otimes \sum_{i=1}^\infty v_kP_k,
\end{equation}
where $\sigma_q^i$ ($i=x,y,z$) is the Pauli operators of the ancilla qubit. Then the quantum process of a RIM sequence of the ancilla is described by a quantum channel, which can be represented in Stinespring representation
\begin{equation}
        \Phi_i(\rho)={\rm Tr}_q[U_i(t_i)(\rho_q\otimes \rho)U_i^{\dagger}(t_i)],
\end{equation}
where ${\rm Tr}_q$ denotes the partial trace over the ancilla, and $U_i=e^{-i Ht_i}=\sum_{\s=0,1}|\s\rangle_q\langle\s|\otimes U_{\s,i}$ with $t_i=2^{m-i}\pi/\kappa$, $U_{\s,i}=e^{-i(-1)^{\s}Vt_i}$, $\rho_q=R_{0}(\pi/2)|\psi\rangle_q\langle \psi|R_{0}^\dagger(\pi/2)$. This quantum channel can also be transformed to the Kraus representation as \cite{Kraus1983,Ma2023c}

\begin{equation}\label{Eq:KrausCell}
    \Phi_i(\rho)=\sum_{\a_i=0}^1 \mathcal{M}_{\a_i}(\rho)=\sum_{\a_i=0}^1M_{\a_i}\rho M_{\a_i}^{\dagger},
\end{equation}
in which $\mathcal{M}_{\a_i}(\cdot)=M_{\a_i} (\cdot) M_{\a_i}^{\dagger}$ is a superoperator with the Kraus operator $M_{\a_i}=[U_{0,i}-(-1)^{\a_i}e^{i\phi_i}U_{1,i}]/2$ and $\phi_{i}=\pi-2\pi0.0\a_{i-1}\dots \a_2 \a_1$ . 

Then the iterative RIMs induce a sequential channel
\begin{equation}
   \Phi_m\cdots\Phi_2\Phi_1(\rho)=\sum_{\a_1,\cdots,\a_m}\M_{\a_m}\cdots \M_{\a_2}\M_{\a_1}(\rho),
\end{equation} 
and for the specific sequence of measurement outcomes $\{\a_1,\cdots,\a_m\}$ representing a quantum trajectory, the state of the target system is steered to
\begin{equation}
     \rho(\a)=\frac{{\M_\a(\rho)}}{{p(\a)}},
\end{equation}
where $\M_\a(\cdot):=\M_{\a_m}\cdots \M_{\a_2}\M_{\a_1}(\cdot)$, $\a:=0.\a_m...\a_2\a_1$ and $p(\a)=p(\a_1,\a_2,\dots,\a_m)=\Tr[\M_\a(\rho)]$ is the probability for this trajectory.

We can prove that \cite{Jin2025}
\begin{equation}\label{Eq:MMa}
\begin{aligned}
	  	     \M_\a(\rho)&=\sum_{k,l=1}^\infty \frac{\sin \left[(\a-v_k/\kappa) 2^{m}\pi\right]}{2^m\sin \left[(\a-v_k/\kappa) \pi\right]}\frac{\sin \left[(\a-v_l/\kappa) 2^{m}\pi\right]}{2^m\sin \left[(\a-v_l/\kappa) \pi\right]}\Pi_k \rho \Pi_l\\
	  	     &=\sum_{k,l=1}^\infty(-1)^{\lfloor \xi_k\rfloor+\lfloor 2^{m}\xi_k\rfloor+\lfloor \xi_l\rfloor+\lfloor 2^{m}\xi_l\rfloor}\sqrt{F_{2^m}(\a-v_k/\kappa)F_{2^m}(\a-v_l/\kappa)}\Pi_k \rho \Pi_l,
\end{aligned}
\end{equation}
where $F_{N}(x):=\left[\frac{\sin(N\pi x)}{N\sin(\pi x)}\right]^2$ is a distorted Fej\'{e}r kernel, and $\xi_k=\a-v_k/\kappa$. Besides, when $2^m$ is relatively large, then the peaks of the Fej\'{e}r kernel are narrow. Only when $\xi_k$ and $\xi_l$ are simultaneously close to integers, the coefficients are finite, and
\begin{equation}
\begin{aligned}
	    &\frac{\sin \left[(\a-v_k/\kappa) 2^{m}\pi\right]}{2^m\sin \left[(\a-v_k/\kappa) \pi\right]}=\frac{\sin \left[(\lfloor\xi_k+1/2\rfloor+\delta) 2^{m}\pi\right]}{2^m\sin \left[(\lfloor\xi_k+1/2\rfloor+\delta) \pi\right]}=\frac{\sin \left(\delta 2^{m}\pi\right)}{(-1)^{\lfloor\xi_k+1/2\rfloor}2^m\sin \left(\delta \pi\right)}=(-1)^{\lfloor\xi_k+1/2\rfloor}\sqrt{F_{2^m}(\a-v_k/\kappa)}.	\\
\end{aligned}
\end{equation}
Here we note that $\lfloor\xi_k+1/2\rfloor=\operatorname{round}(\xi_k)$ and $\delta\to 0$ is a small remainder, then
\begin{equation}
 \M_\a(\rho)\approx \sum_{k,l=1}^\infty(-1)^{\lfloor\xi_k+1/2\rfloor+\lfloor\xi_l+1/2\rfloor}\sqrt{F_{2^m}(\a-v_k/\kappa)F_{2^m}(\a-v_l/\kappa)}\Pi_k \rho \Pi_l,
\end{equation}

By tracing over $\M_\a$, we can obtain the probability to get the measurement outcomes $\a$:
\begin{equation}\label{Eq:Prob.Adap}
    p(\a)=\Tr(\mathcal{M}_\a(\rho))\approx \sum_{k=1}^\infty {\rm Tr}(P_k\rho)F_{2^{m}}(\a-v_k/\kappa).
\end{equation}
The above formula represents a summation of at most $s$ distribution peaks around the set of eigenvalues $\{v_k\}$ with the weight of the $k$th distribution being $\Tr(\rho P_k)$. The peaks are determined by Fej\'{e}r kernels with the half-width being $2^{-m}$, which indicates the error $\delta\sim O(2^{-m})$ for the total time of operation $t=\sum_{i=1}^m 2^{m-i}\pi=(2^m-1)\pi$.

We then consider the case with a free commuting Hamiltonian on the target system, the total Hamiltonian becomes
\begin{equation}
  H=\sigma_q^z\otimes V+\mathbb{I}_q\otimes C,
\end{equation}
where $\mathbb{I}_q$ is the identity operator of the ancilla qubit, and the free Hamiltonian $C$ commutes with $V$, that is, $[V,C]=0$. Then they can be diagonalized simultaneously, giving $C=\sum_{k}c_kP_k$, and $P_k$ is the joint projector of $V$ and $C$. Then $\tu_{\a,i}=U_{\a,i}e^{-iCt_i}$ and $\tm_{\a,i}=M_{\a,i}e^{-iCt_i}$, we have
\begin{equation}
  \begin{aligned}
	  	     \tilde\M_\a(\rho)&=\sum_{k,l=1}^\infty \frac{\sin \left[(\a-v_k/\kappa) 2^{m}\pi\right]}{2^m\sin \left[(\a-v_k/\kappa) \pi\right]}\frac{\sin \left[(\a-v_l/\kappa) 2^{m}\pi\right]}{2^m\sin \left[(\a-v_l/\kappa) \pi\right]}e^{ -i\pi(c_k-c_l) (2^m-1)/\kappa} \Pi_k \rho \Pi_l \\
	  	     &=\sum_{k,l=1}^\infty(-1)^{\lfloor \xi_k\rfloor+\lfloor 2^{m}\xi_k\rfloor+\lfloor \xi_l\rfloor+\lfloor 2^{m}\xi_l\rfloor}\sqrt{F_{2^m}(\a-v_k/\kappa)F_{2^m}(\a-v_l/\kappa)}e^{ -i\pi(c_k-c_l) (2^m-1)/\kappa}\Pi_k \rho \Pi_l.
\end{aligned}
\end{equation}
Specifically, if $C=-V$ is given, or 
$  H=-\ket{1}_q\bra{1}\otimes 2V$, 
\begin{equation}
  \tilde\M_\a(\rho)=  \sum_{k,l=1}^\infty(-1)^{\lfloor \xi_k\rfloor+\lfloor 2^{m}\xi_k\rfloor+\lfloor \xi_l\rfloor+\lfloor 2^{m}\xi_l\rfloor}\sqrt{F_{2^m}(\a-v_k/\kappa)F_{2^m}(\a-v_l/\kappa)} e^{ i\pi(v_k-v_l) (2^m-1)/\kappa}\Pi_k \rho \Pi_l,
\end{equation}
and for $\xi_k,\xi_l$ that near integers, 
\begin{equation}
\begin{aligned}
	    \tilde\M_\a(\rho)&\approx \sum_{k,l=1}^\infty (-1)^{\lfloor\xi_k+1/2\rfloor+\lfloor\xi_l+1/2\rfloor}e^{i\pi (2^m-1)(\lfloor\xi_k+1/2\rfloor-\lfloor\xi_l+1/2\rfloor)}\sqrt{F_{2^m}(\a-v_k/\kappa)F_{2^m}(\a-v_l/\kappa)} \Pi_k \rho \Pi_l	\\
	    &=\sum_{k,l=1}^\infty\sqrt{F_{2^m}(\a-v_k/\kappa)F_{2^m}(\a-v_l/\kappa)} \Pi_k \rho \Pi_l.
\end{aligned}
\end{equation}

\subsection{Measurement statistics of QPE for error detection of rotating symmetric codes}
When we consider QPE for error detection of rotating symmetric codes, the coupling Hamiltonian is
\begin{equation}
  H=-\chi|1\rangle_q\langle 1|\otimes \nn=-\chi|1\rangle_q\langle 1|\otimes \sum_{l=0}^N\sum_{k=0}^\infty (l+kN)\ket{l+kN}\bra{l+kN},
\end{equation}
where $\nn=a^\dagger a$ is the bosonic mode number operator with $a^\dagger (a)$ being the create (destroy) operator and $\chi$ is the dispersive coupling strength. So by choosing $t_i=2\times2^{m-i}\pi/\chi N$, we have
\begin{equation}
    p(\a)=\sum_{l=0}^N\sum_{k=0}^\infty\Tr(\Pi_{l+kN}\rho)F_{2^m}(\a-l/N+k).
\end{equation}
Since $F_N(x)=F_N(x+1)$, we have
\begin{equation}
    p(\a)=\sum_{l=0}^N\sum_{k=0}^\infty\Tr(\Pi_{l+kN}\rho)F_{2^m}(\a-l/N)=\sum_{l=0}^N\Tr(\Pi_N^l\rho)F_{2^m}(\a-l/N),
\end{equation}
where $\Pi_{N}^l=\sum_{k=0}^\infty \ket{l+kN}\bra{l+kN}$.
\subsection{Method and measurement statistics of QPE for error detection of GKP codes}
\begin{figure}[H]
\centering
  \includegraphics[width=15cm]{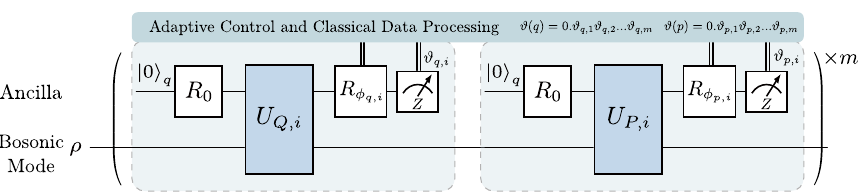}
  \caption{The quantum circuit for detecting displacement error for GKP code. Each cell contains two rounds of RIMs with Hamiltonian $H_Q$ and $H_P$, feedback phase $\phi_{q,i}$ and $\phi_{p,i}$, and evolution time $t_i=\frac{2^{m-i}\pi}{g\sqrt{2\pi}}$ ($i=1,\ 2,\ ...,\ m$).}
  \label{GKPCircuit}
\end{figure}

There are two stabilizers for GKP codes, so we should measure $Q$ and $P$ simultaneously. Here we use the following quantum circuit (Fig. \ref{GKPCircuit}). In each cell, the system is evolved with $U_{Q,i}=e^{-iH_Q t_i}$, followed by $U_{P,i}=e^{-iH_P t_i}$ with $t_i=\frac{2^{m-i}\pi}{g\sqrt{2\pi}}$. The feedback phase for measurements of $Q$ and $P$ are processed separately, i.e. $\phi_{q,i}=\pi-2\pi0.0\a_{q,i-1}...\a_{q,2}\a_{q,1}$ and $\phi_{p,i}=\pi-2\pi0.0\a_{p,i-1}...\a_{p,2}\a_{p,1}$. 

The Hamiltonian of QPE for GKP codes is 
\begin{equation}
\begin{aligned}
      H_Q =\sqrt{2}g\sigma_q^z\otimes Q&=\sqrt{2}g\sigma_q^z\otimes\int_{-\infty}^\infty \left( x\ket{x}_Q\bra{x}\right)\dd x\\
   &=\sqrt{2}g\sigma_q^z\otimes\sum_{k=-\infty}^\infty\int_{-\sqrt{\pi}/2}^{\sqrt{\pi}/2}\left(x+k\pi\right)\ket{x+k\sqrt\pi}_Q\bra{x+k\sqrt\pi}\dd x.
\end{aligned}
\end{equation}
Taking the evolution time $t_i=\frac{2^{m-i}\pi}{g\sqrt{2\pi}}$, we have
\begin{equation}
\begin{aligned}
         p(\a)&=\sum_{k=-\infty}^{\infty}\int_{-\sqrt{\pi}/2}^{\sqrt{\pi}/2} \Tr(\ket{x+k\sqrt\pi}_Q\bra{x+k\sqrt\pi}\rho)F_{2^m}\left(\a-\frac{x}{\sqrt{\pi}}+k\right)\dd x \\
         &=\int_{-\sqrt{\pi}/2}^{\sqrt{\pi}/2}\Tr(\Pi_Q^{x}\rho)F_{2^m}\left(\a-\frac{x}{\sqrt{\pi}}\right)\dd x,
\end{aligned}
\end{equation}
in which $\Pi_{Q}^x=\sum_{k=-\infty}^\infty|x+k\sqrt{\pi}\rangle_Q\langle x+k\sqrt{\pi}|$. 
The resulting state (omitting normalization factor $p[\a(x)]$) turns to
\begin{equation}
    \rho(\a)\propto\sum_{k,l}^{}\int_{-\sqrt{\pi}/2}^{\sqrt{\pi}/2}\dd x_2\int_{-\sqrt{\pi}/2}^{\sqrt{\pi}/2}\dd x_1 (-1)^{b_{x_1,x_2}}\sqrt{F_{2^m}\left(\a-\frac{x_1}{\sqrt{\pi}}\right)F_{2^m}\left(\a-\frac{x_2}{\sqrt{\pi}}\right)}P_Q^{x_1+k\sqrt{\pi}} \rho P_Q^{x_2+l\sqrt{\pi}},
\end{equation}
in which $b_{x_1,x_2}=\lfloor \xi_1\rfloor+\lfloor 2^{m}\xi_1\rfloor+\lfloor \xi_2\rfloor+\lfloor 2^{m}\xi_2\rfloor$, $\xi_i=\a-x_i/\sqrt{\pi}$, and $P_Q^x=|x\rangle_Q\bra{x}$. It is worth noting that when $N$ is large, then the Fej\'er kernel can be considered as a summation of Dirac $\delta-$function, i.e., $F_{N}(x)=\sum_{k=-\infty}^\infty\delta(x-k)$ ($N\to \infty$). Thus when the measurement rounds $m$ is large (then $N=2^m$ is large), we have
\begin{equation}
    \rho(\a)\xrightarrow[]{m\to \infty}\frac{\Pi_Q^{\a\sqrt{\pi }}\rho\Pi_Q^{\a\sqrt{\pi }}}{\Tr(\Pi_Q^{\a\sqrt{\pi }}\rho)},
\end{equation}
which extracts the component of the state projected onto the displaced eigenspaces corresponding to the modular position value.
Similarly,
\begin{equation}
\begin{aligned}
         p(\a)=\int_{-\sqrt{\pi}/2}^{\sqrt{\pi}/2}\Tr(\Pi_P^{p}\rho)F_{2^m}\left(\a-\frac{p}{\sqrt{\pi}}\right)\dd p,
\end{aligned}
\end{equation}
in which $\Pi_{P}^p=\sum_{k=-\infty}^\infty|p+k\sqrt{\pi}\rangle_P\langle p+k\sqrt{\pi}|$ and we note that $[\Pi_Q^q,\Pi_P^p]=0$. The resulting state becomes
\begin{equation}
    \rho(\a)\propto\sum_{k,l}^{}\int_{-\sqrt{\pi}/2}^{\sqrt{\pi}/2}\dd p_2\int_{-\sqrt{\pi}/2}^{\sqrt{\pi}/2}\dd p_1 (-1)^{b_{p_1,p_2}}\sqrt{F_{2^m}\left(\a-\frac{p_1}{\sqrt{\pi}}\right)F_{2^m}\left(\a-\frac{p_2}{\sqrt{\pi}}\right)}P_p^{p_1+k\sqrt{\pi}} \rho P_p^{p_2+l\sqrt{\pi}},
\end{equation}
where $b_{p_1,p_2}=\lfloor \xi'_1\rfloor+\lfloor 2^{m}\xi'_1\rfloor+\lfloor \xi'_2\rfloor+\lfloor 2^{m}\xi'_2\rfloor$, $\xi'_i=\vartheta(p)-p_i/\sqrt{\pi}$, and $P_P^p=|p\rangle_P\bra{p}$. As in the position measurements, when $m$ is large, the state can be simplified to 
\begin{equation}
     \rho(\a)\xrightarrow[]{m\to \infty}\frac{\Pi_P^{\a\sqrt{\pi }}\rho\Pi_P^{\a\sqrt{\pi }}}{\Tr(\Pi_P^{\a\sqrt{\pi }}\rho)}.
\end{equation}
We can also consider the 2-dimensional distribution in the phase space, with
\begin{equation}
\begin{aligned}
         p[\a(x,p)]=\int_{-\sqrt{\pi}/2}^{\sqrt{\pi}/2}\dd p\int_{-\sqrt{\pi}/2}^{\sqrt{\pi}/2}\dd x\Tr(\Pi_Q^{x}\Pi_P^{p}\rho)F_{2^m}\left(\a(x)-\frac{x}{\sqrt{\pi}}\right)F_{2^m}\left(\a(p)-\frac{p}{\sqrt{\pi}}\right).
\end{aligned}
\end{equation}
\section{Example of error detection for binomial codes}
\begin{figure}[htbp]
    \centering
    \includegraphics[width=16cm]{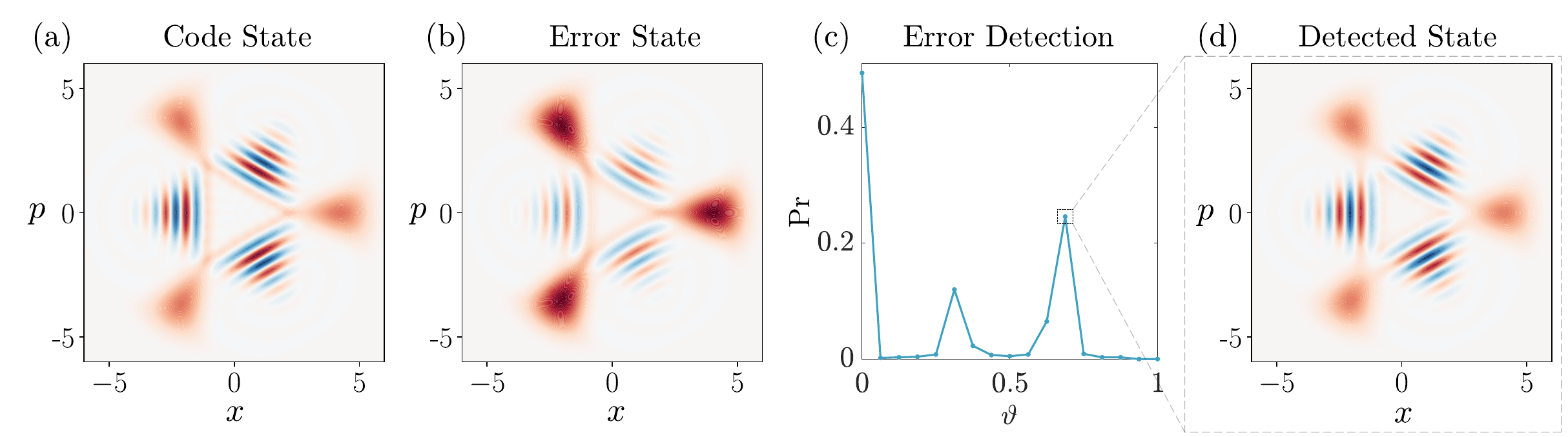}
    \caption{Simulation of error detection of binomial code for an ideal binomial code $\ket{+^{\rm bin}_{3,6}}$ after suffering the lossy error with $\chi=0.1$. The detected state is the quantum trajectory of $\vartheta=0.6875$ which is approximately $\rho'=a\ket{+^{\rm bin}_{3,6}}\bra{+^{\rm bin}_{3,6}}a^\dagger/\Tr(a\ket{+^{\rm bin}_{3,6}}\bra{+^{\rm bin}_{3,6}}a^\dagger)$ with fidelity $F=0.981$.}
    \label{fig:SimBin}
\end{figure}

Binomial code is a class of code in the family of rotational-symmetric bosonic code with parameters $N$ and $K$. The code word is
\begin{equation}
    \ket{\mu_{N,K}^\mathrm{bin}}=\sum_{k=0}^{\lfloor K/2\rfloor-\mu}\sqrt{\frac{1}{2^{K-1}}\begin{pmatrix}
K\\2k+\mu
\end{pmatrix}}\ket{(2k+\mu)N},
\end{equation}
where $\mu\in\{0,1\}$ and $\binom{N}{k}= \frac{N!}{k!(N-k)!}$ represents the binomial coefficient "N choose k". Here we conduct numerical simulation of error detection of binomial code
$\ket{+^{\rm bin}_{3,6}}=(\ket{0^{\rm bin}_{3,6}}+\ket{1^{\rm bin}_{3,6}})/\sqrt{2}$, as shown in Fig. \ref{fig:SimBin}.

\section{Details of finite-energy GKP code}
Here we briefly introduce the basics of finite-energy GKP code \cite{Royer2020}.

The ideal GKP state is defined as an infinite superposition of position operator eigenstates,
\begin{equation}
  \ket{\m^{\rm GKP}}=\sum_{k=-\infty}^\infty \ket{(2k+\m)\sqrt{\pi}}_Q,
\end{equation}
in which $\ket{\cdot}_{Q}$ represents the eigenstate of position operator $Q=(a+a^\dagger)/\sqrt{2}$. The ideal GKP state possesses infinite energy and is non-normalizable, making it physically unrealizable. So the finite-energy GKP states have been developed by introducing a Gaussian envelope that suppresses contributions from high-energy components:
\begin{equation}
  \ket{\mu^{\rm GKP}_\Delta}=\mathcal{N}_\mu e^{-\Delta^2 a^\dagger a}\ket{\m^{\rm GKP}}=\mathcal{N}_\mu \sum_{k=-\infty}^\infty e^{-\Delta^2 a^\dagger a}\ket{(2k+\m)\sqrt{\pi}}_Q.
\end{equation}
Here $\Delta^{-1}$ characterizes the width of the Gaussian envelope, and $\mathcal{N}_\mu$ is the normalization coefficient ensuring $\braket{\mu^{\rm GKP}_\Delta}{\mu^{\rm GKP}_\Delta} = 1$. 

The component states of ideal GKP state suffice $Q\ket{(2k+\m)\sqrt{\pi}}_Q={(2k+\m)\sqrt{\pi}}\ket{(2k+\m)\sqrt{\pi}}_Q$, then for finite-energy GKP $\ket{(2k+\m)\sqrt{\pi}}_{Q_\Delta}=e^{-\Delta^2 a^\dagger a}\ket{(2k+\m)\sqrt{\pi}}_Q$, we have
\begin{equation}
   e^{-\Delta^2 a^\dagger a}Q e^{\Delta^2 a^\dagger a}e^{-\Delta^2 a^\dagger a}\ket{(2k+\m)\sqrt{\pi}}_Q=(2k+\m)\ket{(2k+\m)\sqrt{\pi}}_{Q_\Delta},
\end{equation}
which means $\ket{(2k+\m)\sqrt{\pi}}_{Q_\Delta}$ is the eigenstate of $Q_\Delta=e^{-\Delta^2 a^\dagger a}Q e^{\Delta^2 a^\dagger a}$. Here we note that $Q_\Delta$ is pseudo-Hermitian \cite{Mostafazadeh2005,Scolarici2006,Ohlsson2021}\footnote{The operator $H$ is called $\eta-$pseudo-Hermitian, if there exists an Hermitian and invertible operator $\eta$ satisfying $H^\dagger=\eta H \eta^{-1}$. Pseudo-Hermitian operators have real spectrum when $\eta$ is positive-defined although they are not Hermitian \cite{Mostafazadeh2002}.} since $Q_\Delta^\dagger=e^{\Delta^2 a^\dagger a}Q e^{-\Delta^2 a^\dagger a}=\eta Q_\Delta \eta^{-1}$ with Hermitian, positive and invertible metric operator $\eta=e^{2\Delta^2 a^\dagger a}$. Hence, the spectrum of $Q_\Delta$ is real \cite{Mostafazadeh2002}, which has the physical meaning. Then, the finite-energy GKP state is the $+1$ eigenstate of its stabilizer generated by $Q_\Delta$
\begin{equation}
  S_{z,\Delta}=e^{i2\sqrt{\pi}Q_\Delta}.
\end{equation}

\section{QPE for code preparation}

\subsection{Rotation-symmetric codes}
It is straightforward that our method can be used to construct rotation-symmetric bosonic code. It has been shown that the code state is produced by applying the projector of some subspaces on a primitive state \cite{Grimsmo2020}
\begin{equation}
  |\m^{\rm code}\rangle\propto \Pi_{2N}^{N\m}\ket{\Theta^{\rm code}},
\end{equation}
in which the projector $\Pi_{2N}^{N\m}=\sum_{k=0}^\infty \Pi_{2kN+N\m}$, and $\ket{\Theta^{\rm code}}$ is the primitive state specified for different codes. For example, the primitive state of cat code is $\ket{\Theta^{\rm cat}_{\alpha}}=\ket{\alpha}$. Then the density matrix of the code state is
\begin{equation}
\begin{aligned}
	    \rho_\m^{\rm code}&\propto\Pi_{2N}^{N\m}\ket{\Theta^{\rm code}}\bra{\Theta^{\rm code}}\Pi_{2N}^{N\m}	\\
	    &=\sum_{k,l=0}^{\infty}\Pi_{2kN+N\m}\ket{\Theta^{\rm code}}\bra{\Theta^{\rm code}}\Pi_{2lN+N\m}.
\end{aligned}
\end{equation}
Then we consider the state under the superoperator of sequential RIM evolving with dispersive coupling and evolution time $t_i=2\times2^{m-i}\pi/\chi (2N)$,
\begin{equation}
  \tilde\M_\a(\rho)\approx \sum_{k,l=1}^\infty\sqrt{F_{2^m}(\a-k/2N)F_{2^m}(\a-l/2N)} \Pi_{k} \rho \Pi_l.
\end{equation}
So for the outcome $\a=0$
\begin{equation}
\begin{aligned}
	    \M_0(\rho)&\approx \sum_{k,l=1}^\infty\sqrt{F_{2^m}(0-k/2N)F_{2^m}(0-l/2N)} \Pi_{k} \rho \Pi_l	\\
	    &\approx \sum_{k,l=1}^\infty \Pi_{2kN} \rho \Pi_{2lN},
\end{aligned}
\end{equation}
here the approximation $F_{2^m}(0-k/2N)=\delta_{k\mod 2N,0}$ is taken for $2^m$ being relatively large. Likewise, by selecting the outcome $\a=(0.5)_{\rm D}=(0.1)_{\rm B}$ \footnote{$(\cdot)_{\rm B,D}$ represent numbers expressed in binary and decimal, respectively.}, we have
\begin{equation}
  \M_{0.5}(\rho)\approx \sum_{k,l=1}^\infty \Pi_{2kN+N}\, \rho \,\Pi_{2lN+N}.
\end{equation}
So the outcome state with $\theta=(0.\mu)_{\rm B}$ and initial state $\ket{\Theta^{\rm code}}\bra{\Theta^{\rm code}}$ is exactly the code state $\rho_\m^{\rm code}$.

\begin{figure}[H]
\centering
  \includegraphics[width=16cm]{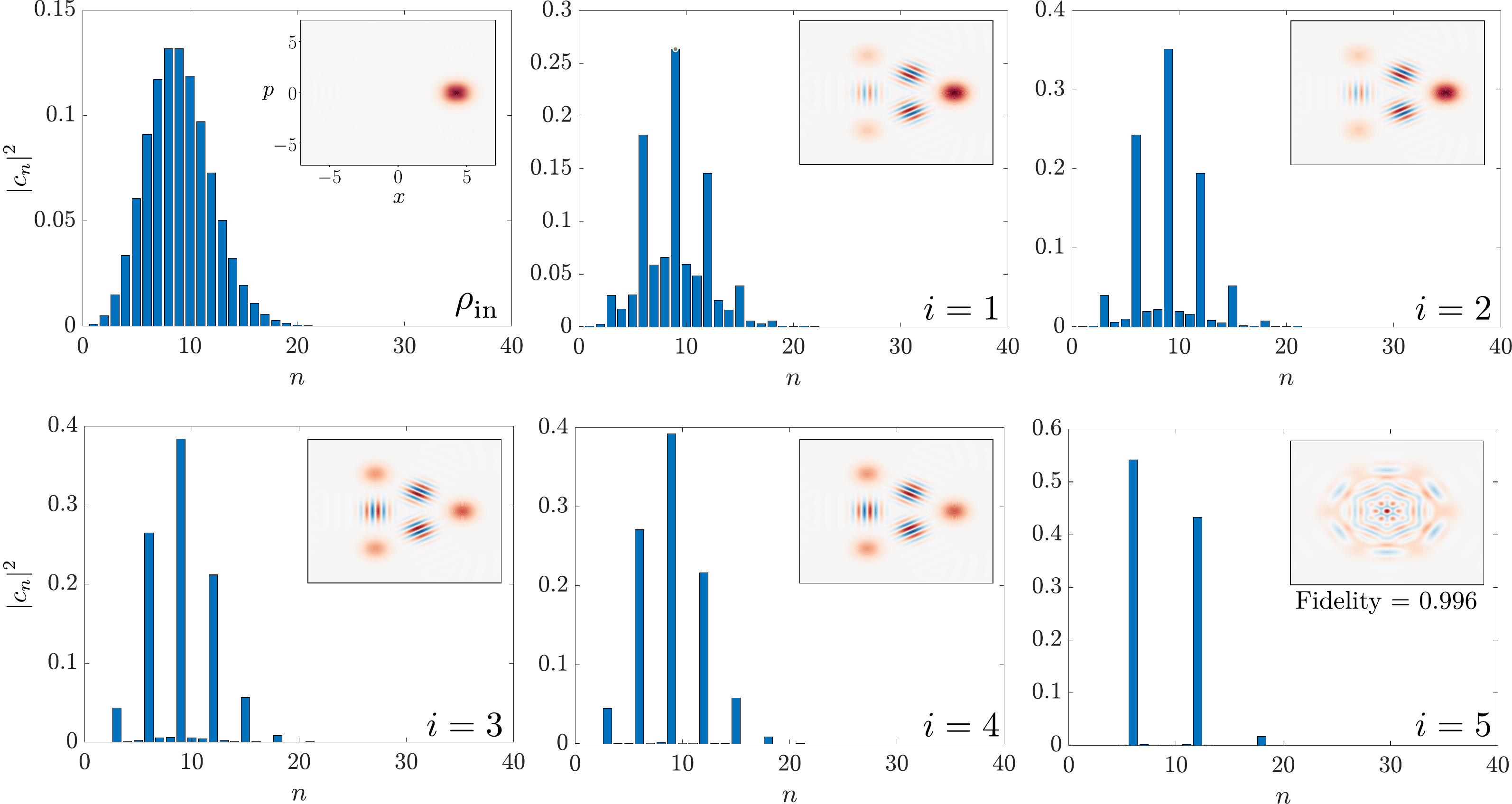}
  \caption{The photon number populations and Wigner functions after each round of RIM in producing $\ket{0^{\rm cat}_{3,3}}$. The input state $\rho_{in}=\ket{\alpha=3}\bra{\alpha=3}$ is a coherent state, then by selecting the trajectory with all outcomes being 0, the cat state is generated.}\label{constructCat}
\end{figure}

\begin{figure}[H]
\centering
  \includegraphics[width=16cm]{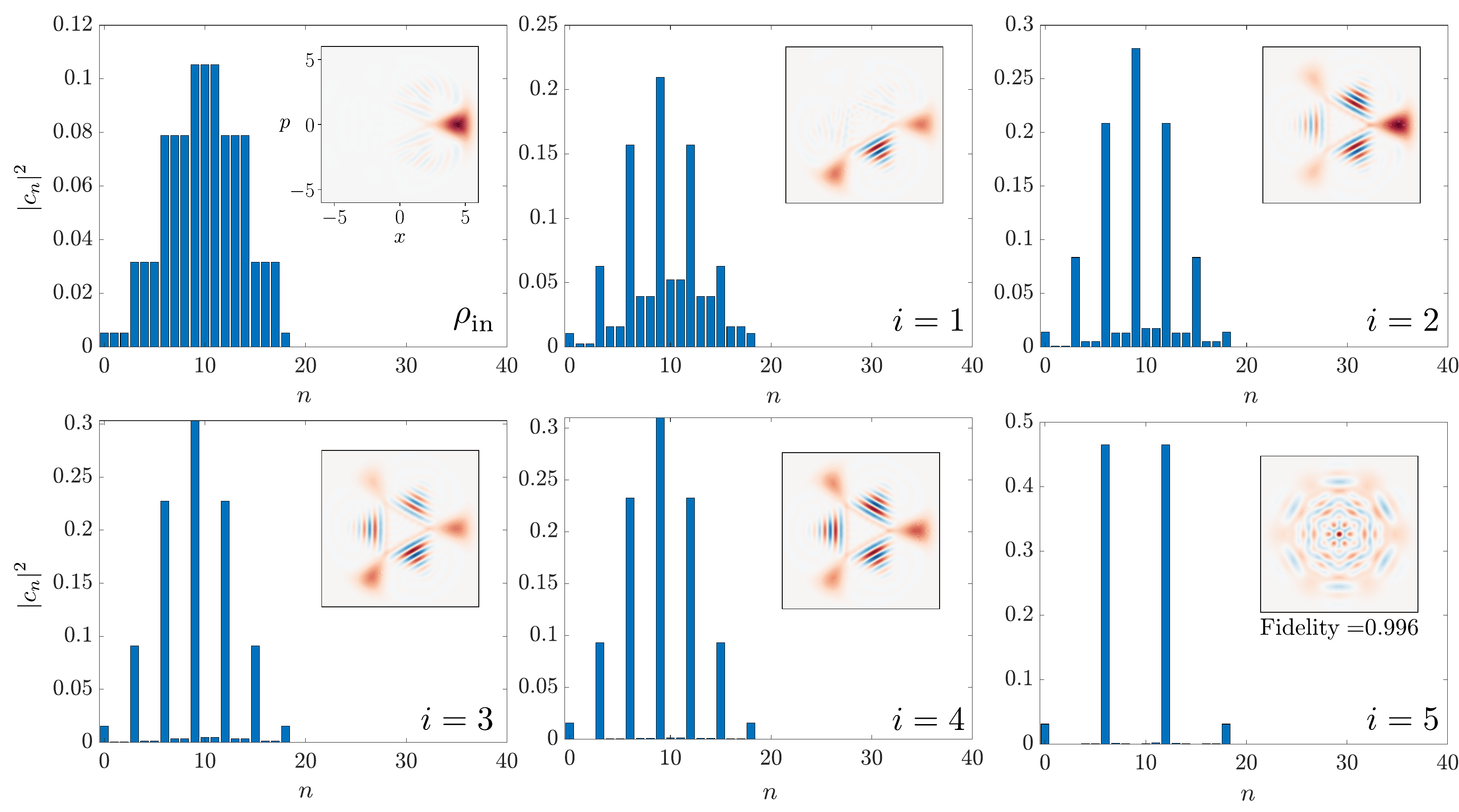}
  \caption{The photon number populations and Wigner functions in each round of RIM in producing $\ket{0^{\rm bin}_{3,6}}$. The input state is $\rho_{in}=\ket{\Theta^{\rm bin}_{3,6}}\bra{\Theta^{\rm bin}_{3,6}}$, then by selecting the trajectory with all outcomes being 0, the cat state is generated.}\label{constructBin}
\end{figure}


\begin{figure}
    \centering
    \includegraphics[width=16cm]{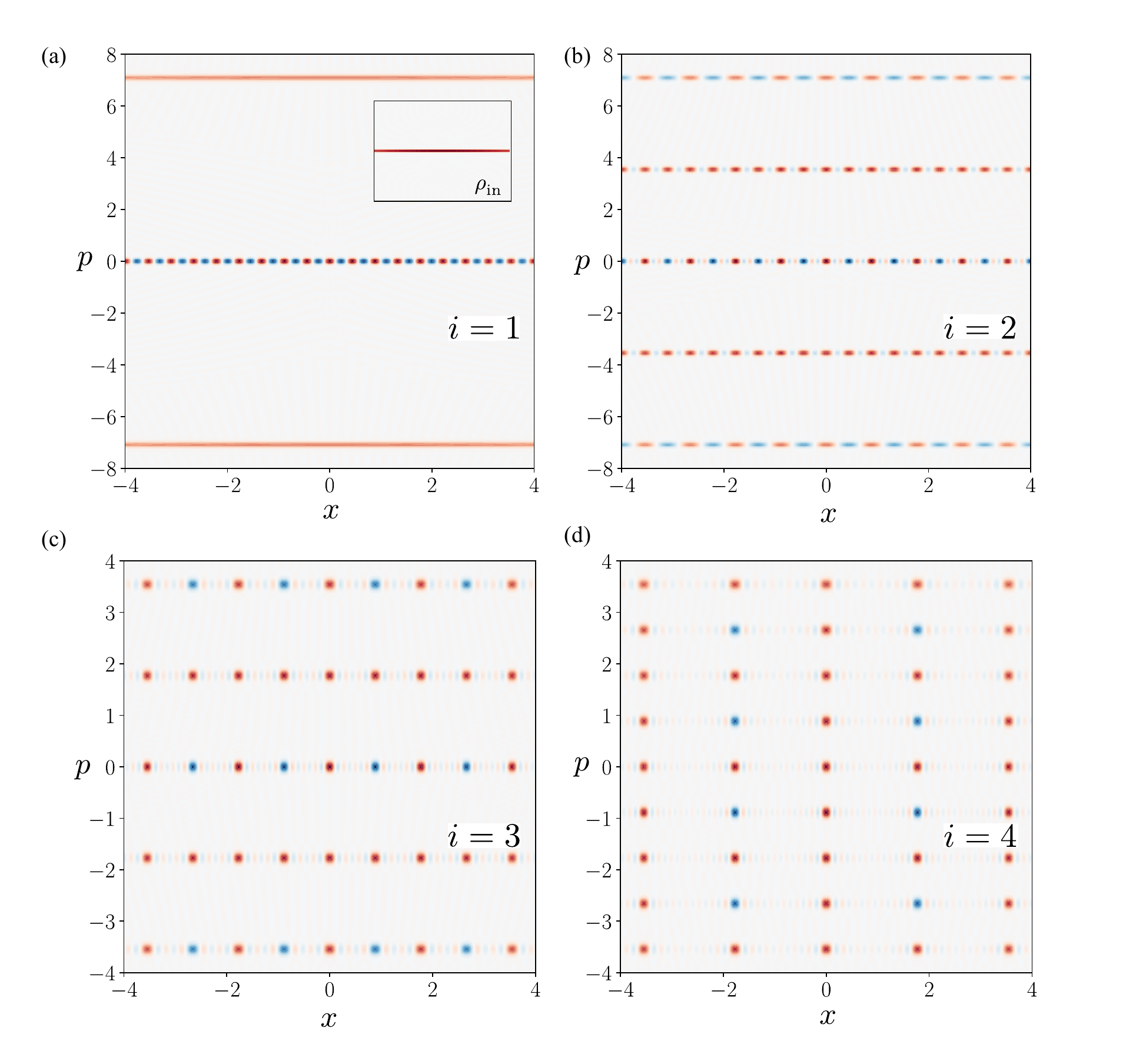}
    \caption{GKP state encoding. The Wigner functions in each cycle of RIM are shown.}
    \label{fig:GKPencode}
\end{figure}

\subsubsection{Cat codes}

For example, Fig. \ref{constructCat} shows the construction of the cat state $|0^{\rm cat}_{3,3}\rangle$, in which the quantum trajectory with $a=0$ is selected from the primitive coherent state
\begin{equation}
    |\Theta^{\rm cat}_{N,\alpha}\rangle=|\alpha\rangle.
\end{equation}
The construction fidelity reaches 0.996.

\subsubsection{Binomial codes}
The binomial code can also be constructed with its primitive state \cite{Grimsmo2020}
\begin{equation}
    \ket{\Theta^{\rm bin}_{3,6}}\propto \sum_{n=0}^{kN}\sqrt{\frac{1}{2^{K-1}}\begin{pmatrix}
K\\\lfloor\frac{n}{N}\rfloor.
\end{pmatrix}}
\end{equation}
Here we show a preparation of $|0_{3,6}^{\rm bin}\rangle$ in Fig. \ref{constructBin} through our QPE approach, and the fidelity also reaches 0.996.

\subsection{GKP code}
Our protocol is also capable for encoding GKP state, the idea is similar to \cite{Terhal2016}, but we incorporate it into the framework of this general approach to interpret.

We start with a squeezed vacuum state $S(r)|0\rangle$ \cite{Hastrup2021a}, in which $S(\zeta)\equiv\exp\left(\left(\zeta^*a^2-\zeta a^{\dagger2}\right)/2\right)$, by taking $r<0$ and $r\in \mathbb{R}$ to obtain a momentum-squeezed state [inset of Fig. \ref{fig:GKPencode} {\color{blue}(a)}]. This state looks like a state that “filled with the eigenstate of $Q$”, so an appropriate selection of  $Q$’s eigenstate can yield the GKP state. Specifically, since the   $|0^{\rm GKP}\rangle $ state is a superposition of $\ket{ q = 2k\sqrt{\pi}}$ , performing QPE on this state and selecting the eigenstate of $ q=0 \mod 2\sqrt{\pi} $ corresponds to the   $|0^{\rm GKP}\rangle $  state. It can be realized by taking the evolution time $t_i=\frac{2^{m-i}\pi}{2g\sqrt{2\pi}}$ and selecting the quantum trajectory with $\a=0$.

\section{Details of arbitrary Fock state generation}

We first introduce our Chinese remainder theorem (CRT) based scheme in detecting (constructing) Fock state.

\textbf{Theorem 1} (\textit{Chinese remainder theorem}) Let \(\{N_i\}_{i=1}^r\) be a set of pairwise coprime integers. Then, for any given integer system of congruences
\begin{align}
\begin{cases}
	  x= l_1\,(\mod\,N_1),\\
	  x= l_2\,(\mod\,N_2),\\
	  \cdots\\
	  x= l_r\,(\mod\,N_r),\\
\end{cases}
\end{align}
there is a unique solution for $x$ modulo $M=\prod_{i=1}^r \nm_i$. Moreover, this solution may be written explicitly as 
 $x=\sum_{i=1}^r l_iM_iy_i$, in which $M_i=M/N_i$ and $y_i$ is the inverse of $M_i$ in modulo $m_i$, i.e. $M_iy_i=1\mod m_i$.
 
 Applying the CRT to our bosonic system allows us to detect large Fock states by performing measurements in several pairwise coprime moduli. Specifically, once one obtains the measurement outcomes $\{l_1, l_2, \dots, l_r\}$ with respect to these different moduli $\{N_i\}$, the unique photon number $n$ is determined via the above CRT formula. Conversely, to construct a large Fock state, one may start with a coherent state that has non-zero overlap with the desired Fock state. By post-selecting the measurement outcomes that project onto the correct modular numbers $\{l_i (\mod\, N_i)\}$ across the different rounds of measurement, one can collapse the coherent state onto the specific large Fock state.

\begin{figure}[htbp]
\centering
  \includegraphics[width=13cm]{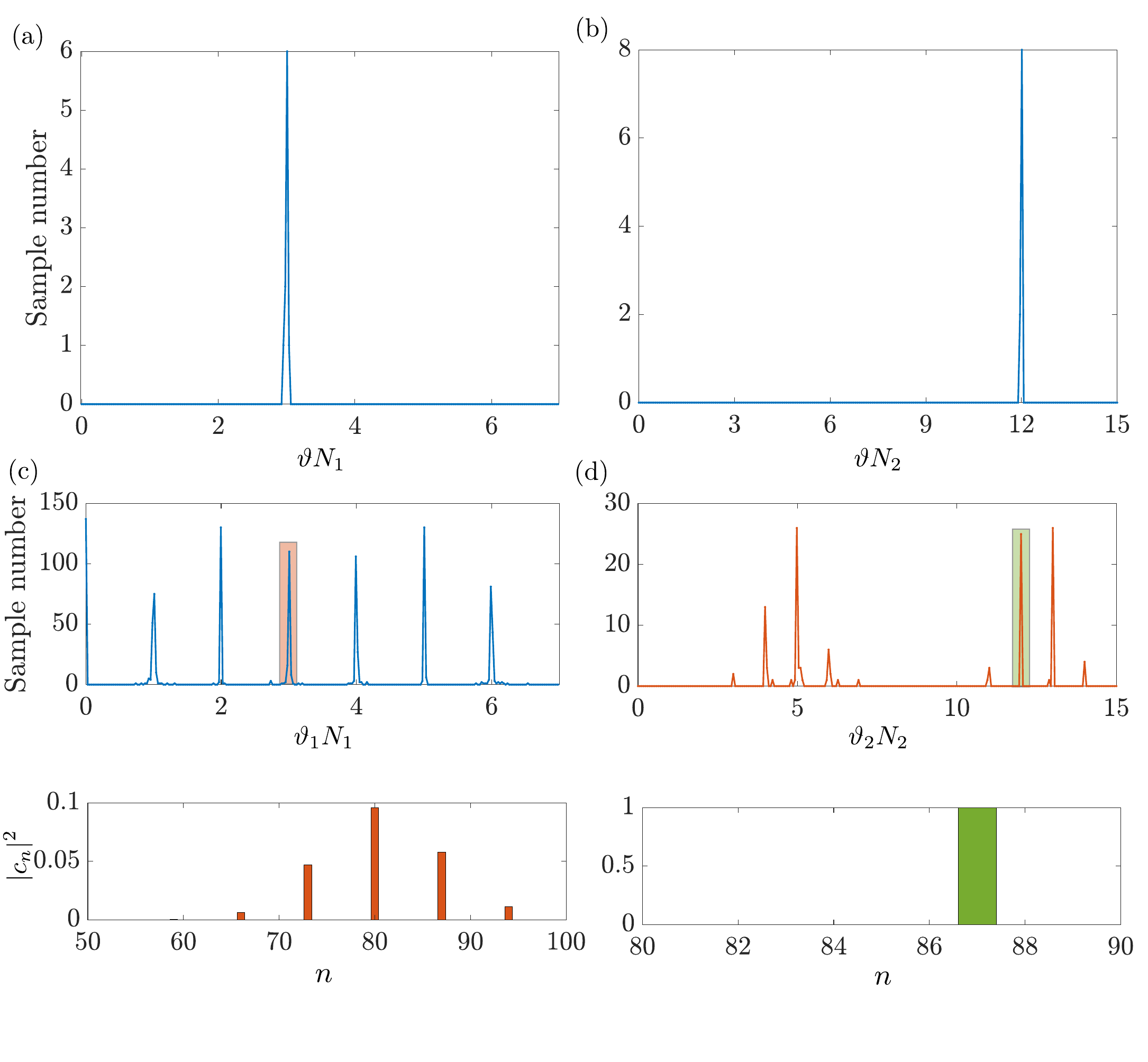}
  \caption{Large Fock state generation. (a-b) The histogram of Monte Carlo simulation for a input state $\ket{87}$ with (a) $\nm=7$ and (b) $\nm=15$. We use $10$ samples and the parameter is $m=8$. We note that the photon number can be determined for a single sample since the peak is quite narrow. (c-d) The histogram of Monte Carlo simulation (upper panels) and photon number populations of post-selected state (lower panels) for a coherent state $\ket{\alpha=9}$ input. Fock state $\ket{87}$ (lower panel of (d)) is gained with setting $\nm_1=7$ and post selecting the state with $\a_1\approx3/7$ (lower panel of (c)) as the input of the second QPE experiment with $\nm_2=15$, followed by post-selecting the state with $\a_2 \approx 12/15$.}\label{Fig:CRT}
\end{figure}

 For example, the photon number of state $\ket{87}$ [$87=3\,(\mod\,7)=12\,(\mod\,15)$] is detected by choosing $\nm_1=7$ and $\nm_2=15$ sequentially [see Fig. \ref{Fig:CRT} {\color{blue}(a-b)}]. Inversely, by post-selecting the trajectories for input state $\ket{\alpha=9}$ with $\a_1\approx3/7$ (as the input state of the second experiment with $\nm_2=15$) and $\a_2\approx 12/15$, the Fock state $\ket{87}$ is constructed [see Fig. \ref{Fig:CRT} {\color{blue}(c-d)}].
 
\section{Performance of error detection}
Here we evaluate the performance of error detection for our method. We mainly consider deduction error ($\delta_D$) caused by the broadening peak with small $m$ ($t_{\rm tot}$) and relaxation infidelity ($\delta_R$) caused by relaxation of the bosonic mode and the ancilla. 

The total infidelity caused by both deduction and relaxation is 
\begin{equation}
    \delta(m,N)=\sum_{l=0}^{N-1}\sum_{\vartheta\in r_{N,l}}p_{\vartheta}[1-F(\rho_{\vartheta},\rho_l)]
\end{equation}
where $r_{N,l}=[l/N-1/2N,l/N+1/2N)$ is the error subspace in which the $N$-fold rotational symmetric code gained (lost) $l$ ($N-l$) photons, $\rho_{\vartheta}=\mathcal{M}_\vartheta(\rho_{\rm in})/p(\vartheta)$ is the output state corresponding to the outcome $\vartheta$, $\rho_l=\Pi_N^l\rho_{\rm in}\Pi_N^l/\Tr(\Pi_N^l\rho_{\rm in}\Pi_N^l)$ is the reference state in the error subspace corresponding to the loss of $l$ photons, and $F(\sigma,\rho)=\left(\Tr\sqrt{\rho^{1/2}\sigma\rho^{1/2}}\right)^2$ is the fidelity between two quantum states $\rho$ and $\sigma$.

\subsection{Deduction errors}
\begin{figure}[htbp]
    \centering
    \includegraphics[width=16cm]{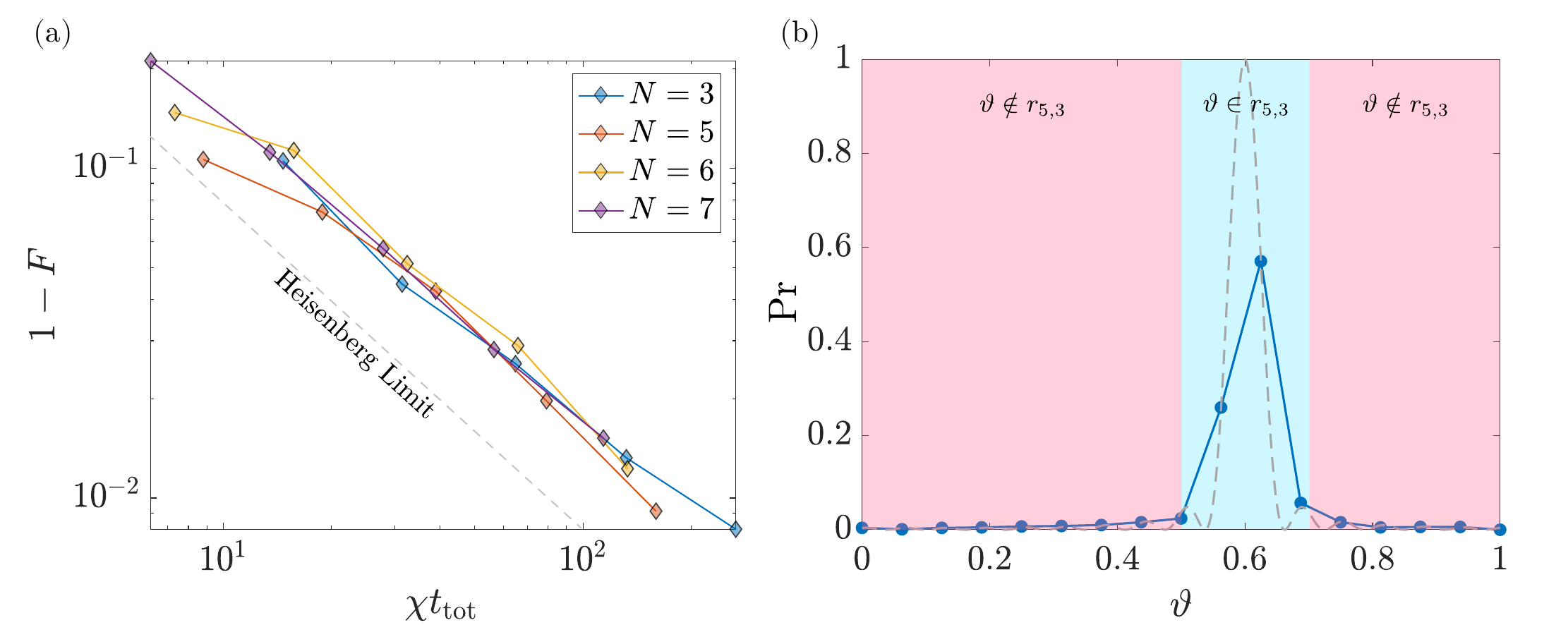}
    \caption{(a) The deduction infidelity $\delta_\mathrm{D}$ as a function of total evolution time $\chi t_{\rm tot}=2\times (2^{m-1}-1)\pi/N$ for different $m$ and $N$. The performance is similar between the modular number $N$ and Heisenberg scaling of error ($\delta\propto t_{\rm tot}^{-1}$) is indicated. The input state is $\rho_{\rm in}=\exp{\xi\mathcal{D}(\rho_+^{\rm{cat}})}$ with $\xi=0.15$. (b) Illustration of the deduction infidelity for $N=5$ and $l=3$ with $m=4$. The histogram shows the detection outcomes, where the region $r_{5,3} = [3/5 - 1/10, 3/5 + 1/10)$, corresponding to detecting two-photon loss errors, is highlighted in blue. Regions outside $r_{5,3}$, corresponding to other error subspaces, are marked in red. Deduction infidelity arises from the distribution of the Fej\'er kernel's peak into incorrect error subspaces. }
    \label{fig:error}
\end{figure}

The statistic distribution shows at most $N$ peaks of Fej\'{e}r kernel around $\a=l/N$ for $l=0,1,..., N-1$. For the outcome $\a\in r_{N,l}$, the bosonic mode is deduced to gain (loss) $l$ ($N-l$) photons.  Then the deduction infidelity arises from the broaden of Fej\'er kernel, i.e. the possibility for Fej\'er kernel around $l/N$ distributing out of $r_{N,l}$, see Fig. \ref{fig:error}{\color{blue}(b)} for an illustration.

The deduction infidelity of different $m$ and $N$ is calculated numerically and shown in Fig. \ref{fig:error}{\color{blue}(a)}. We note that for $N=2^q$, the deduction infidelity is zero since the eigenvalues $l/N$ can be expanded in binary number with finite bits and the error of photon loss is exactly determined with $m\geq q$. 

\subsection{Errors from noisy ancilla and bosonic systems}
Here we consider the performance of error syndrome detection in the practical cQED system with the presence of transmon and cavity relaxation. The parameters in the simulations are listed in Table. \ref{Table}

\begin{table}[H]
\caption{{Parameters for simulations.}}
\centering
\begin{tabular}{cc}
\toprule
Parameter&Value \\
\midrule
Dispersive coupling strength $\chi/2\pi$& $2\mathrm{MHz}$ \\
Longitudinal coupling strength $g/2\pi$& $21.5\mathrm{MHz}$ \\
Ancilla relaxation rate $\Gamma_1$ & $2\times10^{-2}\mathrm{\mu s}^{-1}$\\
Cavity lossy rate $\Gamma_2$ & $10^{-3}\mathrm{\mu s}^{-1}$\\
\bottomrule
\end{tabular}
\label{Table}
\end{table}

The evolution of the whole system can be described by the following Lindblad master equation,
\begin{equation}
  \dv{\rho_{\rm tot}}{t}=-i[H,\rho_{\rm tot}]+\sum_k \Gamma_k \left (L_k\rho_{\rm tot} L_k^{\dagger}-\frac{1}{2}\left\{L_k^{\dagger} L_k,\rho_{\rm tot}\right\}\right),
\end{equation}
where $\rho_{\rm tot}$ is the density matrix of the composite system, $H=-\chi|1\rangle_q\langle 1|\otimes \nn$ is the Hamiltonian of dispersively coupling, $L_1=\sigma_-\otimes \mathbb{I}=|1\rangle\langle0|\otimes \mathbb{I}$ denotes the relaxation of the ancilla, $L_2=\mathbb{I}_q\otimes a$ denotes the relaxation of the bosonic mode and $\Gamma_k$ is the dissipation rate.

 \subsubsection{Rotation-symmetric code}

\begin{figure}[htbp]
\centering
  \includegraphics[width=16cm]{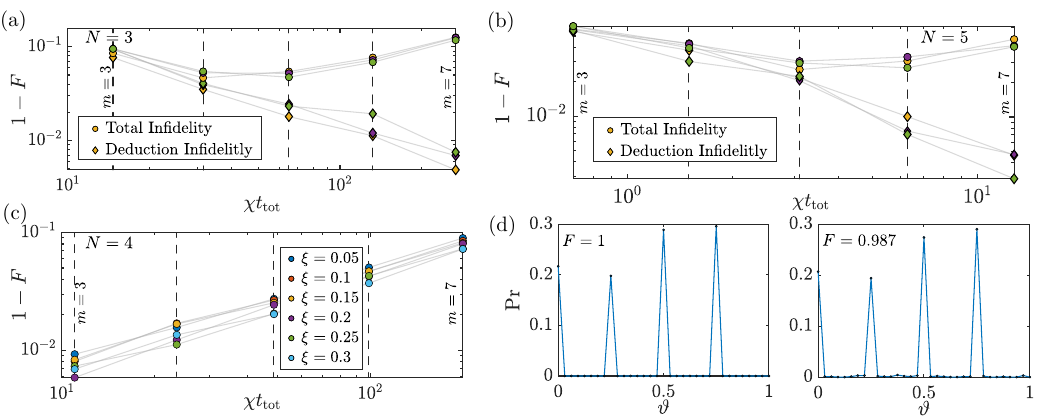}
  \caption{Comparison of infidelity. (a-c) Total infidelity of (a) $N=3$, (b) $N=4$ and (c) $N=5$ cat code (note that the deduction infidelity of $N=4$ rotational code is zero, then total infidelity is caused solely by relaxation) for input state $\rho_{\rm in}=\exp{\xi\mathcal{D}(\rho_+^{\rm{cat}})}$ with various $\xi$.  (d) The probability distribution without (left, no infidelity) and with (right, with relaxation infidelity) relaxation for the input state with $\xi=0.15$ and $m=5$ for $N=4$ cat code are shown in right.}
  \label{fig:infidelitycomp}
\end{figure}

Figure \ref{fig:infidelitycomp} shows the total infidelities of the cat codes with $N=3,4,5$ as functions of the input states and the number of measurements $m$. The relaxation infidelity $\delta_{\rm R}$ is evaluated by measuring total infidelity for $4$-fold rotational code since the deduction infidelity is zero [see the left probability distribution in Fig. \ref{fig:infidelitycomp}{\color{blue} (d)}]. Figure \ref{fig:infidelitycomp}(c) shows that $\delta_{\rm R}$ increases as the total measurement time increases. When including the deduction infidelity, the total infidelity of the $N=3$ and $N=5$ rotational codes first decreases and then increases with $m$. Consequently, an optimal measurement number $m$ exists [see Fig. \ref{fig:infidelitycomp}(a–b)].

\begin{figure}
    \centering
    \includegraphics[width=16cm]{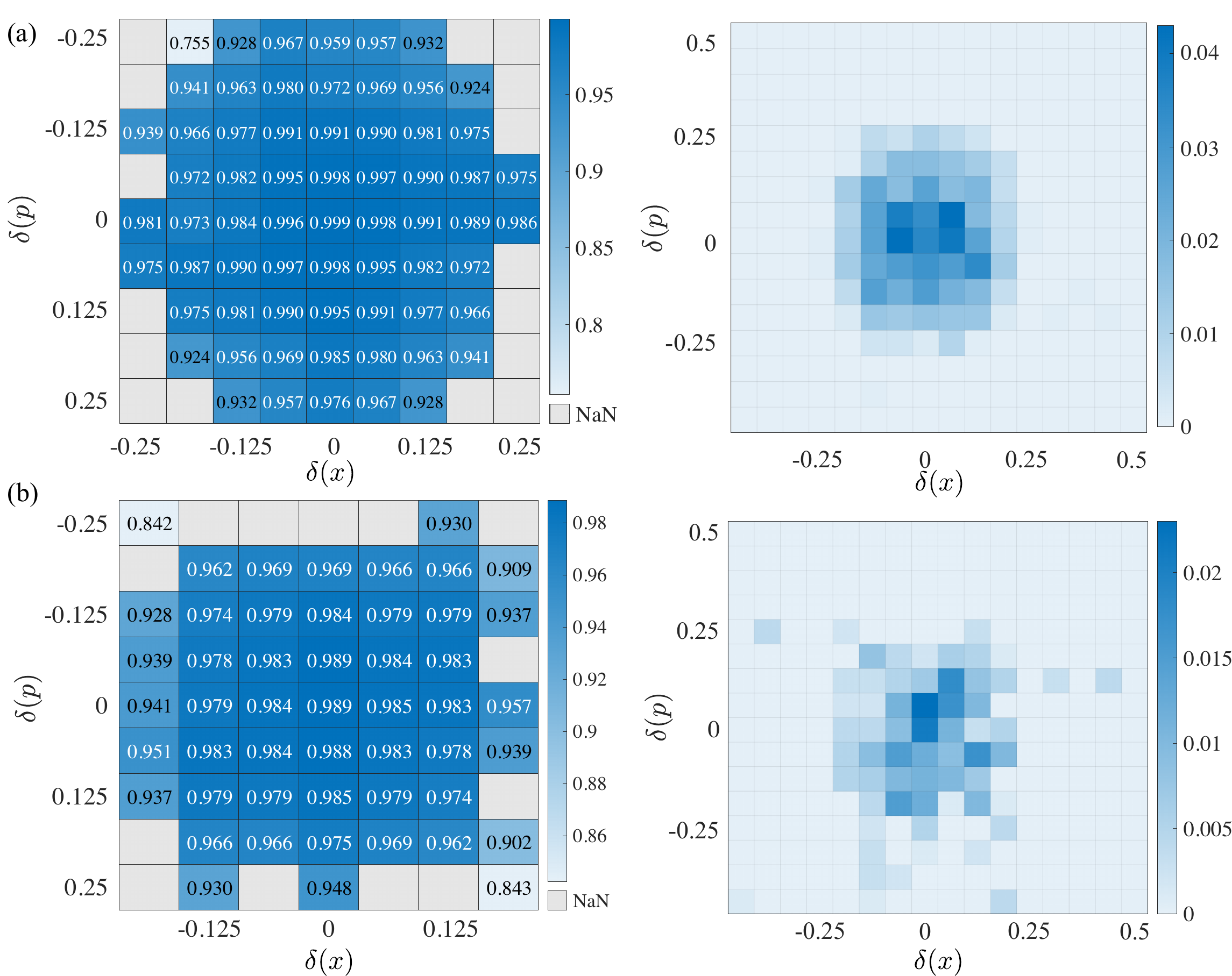}
    \caption{The fidelity of GKP code error detection for all the detection outcomes (a) without noise and (b) with noise. The marginal probability distributions of the detection outcomes of each quadrature are shown in the right.}
    \label{fig:GKPperf}
\end{figure}

\subsubsection{GKP code}
For the GKP code, each trajectory with measurement outcomes $\a(x)$ and $\a(p)$ corresponds to a state $\rho'=D[\a(x),\a(p)]\rho D^\dagger[\a(x),\a(p)]$, in which $D[\a(x),\a(p)]=D[(\delta(x)+i\delta(p))\sqrt{\pi/2}]$, and $\delta(x)=\a(x)-\operatorname{round}[\a(x)]$ represents the actual displacement. Then we analyze the fidelity of each error subspaces with (without) noise [see Fig. \ref{fig:GKPperf}], the average fidelity with (without) error is 0.970 (0.989) with the coupling strength $g=2\pi\times 21.5$MHz \cite{Didier2015}.

\bibliography{Bosonic2}